\documentclass[11pt, letterpaper]{article} % Se define el papel aquí para evitar conflictos

\usepackage[T1]{fontenc}
\usepackage{lmodern}
\usepackage[margin=1in]{geometry} % Cargado SOLO UNA VEZ
\usepackage{amsmath,amssymb}

\usepackage{newtxtext,newtxmath}

\usepackage{graphicx}
\usepackage{booktabs}
\usepackage{threeparttable}
\usepackage{float}
\usepackage{placeins}
\usepackage{caption}

\usepackage{bbm}

\usepackage[numbers,sort&compress]{natbib}
\usepackage{url}

\usepackage{hyperref}
\hypersetup{colorlinks=true,linkcolor=black,citecolor=black,urlcolor=black}

\newcommand{\parencite}[1]{\cite{#1}}
\newcommand{\textcite}[1]{\cite{#1}}

\title{\textbf{Structurally Conditioned Diffusion Reproduces Skills-Based Stratification}}
\author{
Roberto Cantillan$^{1}$ and Mauricio Bucca$^{1}$\\
\small $^{1}$Department of Sociology, Pontificia Universidad Católica de Chile, Santiago, Chile
}
\date{}

\begin{document}
\maketitle

% ======================
% ABSTRACT (≤150 words; no citations)
% ======================
\begin{abstract}
Occupational hierarchies remain strikingly stable even as the content of work changes rapidly. A central missing piece is whether the propagation of skill requirements across occupations is directionally neutral along the wage hierarchy, or instead follows systematic channels that privilege some trajectories over others. Using O*NET (2015--2024), we analyze 17.3 million directed diffusion opportunities linking 873 occupations and 161 skills and show that propagation obeys an \emph{Asymmetric Trajectory Channeling} (ATC) rule: the same requirement spreads differently upward and downward, and these asymmetries depend on skill domain and on the architecture of skill dependencies. ATC is generated by two structural mechanisms. \emph{Directional incorporation asymmetry} implies that wage gradients create distinct receiving environments---upward-moving socio-cognitive requirements encounter complementary infrastructure already in place, whereas upward-moving sensory/physical requirements face structural indifference. \emph{Structural portability constraint} implies that dependency position governs portability---requirements anchoring long prerequisite chains carry co-adoption burdens that restrict diffusion regardless of destination. Consistent with these mechanisms, socio-cognitive requirements propagate upward more often than downward (20.7\% vs.\ 14.9\%), while sensory/physical requirements exhibit the mirror pattern (19.5\% downward vs.\ 10.3\% upward). Moreover, nestedness amplifies these asymmetries in opposite directions: scaffolding capabilities ascend most readily, whereas structurally embedded physical requirements are most tightly confined. Identification leverages within-occupation variation in propagation direction, and results are robust to origin- and destination-side specifications. Together, these findings reveal a directional architecture of occupational change: the diffusion of job requirements is systematically channeled in ways that can reproduce hierarchy through ongoing reconfiguration, even absent assortative preferences or coordinated action.
\end{abstract}

\begin{quote}
\textbf{Significance ---} Why do labor-market hierarchies persist even as job content changes? We identify a simple rule governing how occupational skill requirements spread. Socio-cognitive requirements move upward in the wage hierarchy with little resistance, while sensory and physical requirements tend to remain lateral or shift downward --- not because they are actively rejected, but because high-wage occupations show little responsiveness to them while lower-wage occupations are strongly receptive. This asymmetric channeling reproduces inequality from within the evolution of job requirements and identifies concrete pathways for expanding occupational upgrading.
\end{quote}

\clearpage

\section*{INTRODUCTION}

Occupational inequality persists over time even as the content of jobs continuously changes \parencite{acemoglu_skills_2011, autor_skill_2003}. High-wage occupations differ from low-wage ones not only in pay but in the structure of what they require: their skill portfolios are broader, more cognitively intensive, and more hierarchically organized. Yet both the sources of this hierarchy and the processes that sustain it as job content evolves remain incompletely understood. Whether the structure that sorts skills in space also governs the direction in which they move through it remains an open question. Our central claim is that the skill space is direction-dependent: the same structural proximity entails different propagation frictions for upward versus downward moves along the wage hierarchy.

Recent advances have mapped this structure with increasing precision. Alabdulkareem et al. \parencite{alabdulkareem_unpacking_2018} show that workplace skills form two structurally distinct clusters --- socio-cognitive and sensory/physical --- that predict wages, educational requirements, and automation exposure. Hosseinioun et al. \parencite{hosseinioun_skill_2025} show that within each cluster, skills are hierarchically nested: occupations with narrower portfolios hold subsets of the requirements found in richer ones, and this dependency structure has grown more pronounced over two decades. These accounts are fundamentally architectural: they identify where skills sit in occupational space, not the rules that govern where they move. Because neither framework distinguishes upward from downward propagation, neither can test whether a requirement's structural position conditions the direction it travels --- or whether directional frictions in the propagation process itself reproduce the very hierarchy these accounts describe. Here we address this gap directly. We apply an interval-censored gravity model to rolling-panel data from O*NET (2015--2024) --- spanning approximately 17.3 million diffusion opportunities --- to estimate directional friction parameters separately for upward and downward propagation, and to test how a skill's domain and nestedness position shape those frictions.

We term this propagation rule Asymmetric Trajectory Channeling (ATC). It is generated by two structural mechanisms operating jointly: directional incorporation asymmetry, whereby the wage gradient between source and target creates skill-type--specific incorporation environments---socio-cognitive requirements moving upward encounter complementary infrastructure already in place, while sensory/physical requirements encounter structural indifference, not active exclusion but the absence of organizational conditions that make incorporation coherent; and structural portability constraint, whereby a skill's position in the dependency architecture restricts how freely it can diffuse---requirements anchoring long prerequisite chains carry co-adoption burdens that grow with dependency depth. Intuitively, high-wage occupations can absorb additional socio-cognitive requirements because they already bundle complementary tasks, tools, and organizational routines, whereas the same destinations show little responsiveness to additional sensory/physical requirements. Together, these mechanisms yield a directional rule: socio-cognitive requirements face negligible resistance as they propagate upward in the occupational wage hierarchy, while sensory/physical requirements are disproportionately channeled into lateral or downward trajectories. Nestedness amplifies these asymmetries in opposing directions---scaffolding capabilities ascend most readily, while deeply embedded physical requirements face the steepest confinement---yielding a structured typology of trajectories whose consequences for occupational stratification persist even absent assortative preference or coordinated action: hierarchy is sustained by the directional architecture of the skill space itself as requirements propagate. By specifying where requirement change is most and least likely to travel, ATC provides a structural micro-foundation for why occupational polarization can persist even under continuous task redefinition.

Specifically, we address three questions. \emph{Q1.} Is skill diffusion directional --- do socio-cognitive and sensory/physical requirements face systematically different upward versus downward friction, net of structural distance and occupational heterogeneity? \emph{Q2.} Does nestedness position moderate directional friction in predictable and opposing ways across skill domains? \emph{Q3.} Are these patterns robust across destination-side and origin-side identification strategies, supporting interpretation of ATC as a property of the diffusion channel rather than of either endpoint in isolation? In the next section, we develop the theoretical framework and derive the ATC typology. We then present the gravity model and results, and discuss implications for stratification theory and labor market policy.

\section*{Theory}

\subsection*{The occupational system as a constrained diffusion field}

We conceptualize skill propagation as a gravity process operating over a  structured occupational space. The basic intuition is familiar from models of trade, migration, and information flow \parencite{tinbergen_shaping_1962, hidalgo_product_2007}: the probability that a requirement spreads from one occupation to another increases with the capacity of both occupations and decreases with the structural distance between them. Applied to skill diffusion, this logic yields a parsimonious decomposition:

\begin{equation}
    P_{ijk} \propto \frac{M_i \cdot M_j \cdot S_k(\mathbf{x}_k)}{D_{ij}^{\gamma}}
    \label{eq:gravity}
\end{equation}

\noindent where $M_i$ captures the emission capacity of the source occupation and $M_j$ the absorptive capacity of the target. $S_k(\mathbf{x}_k)$ captures the intrinsic diffusibility of skill $k$ as a function of its structural characteristics, allowing requirements to differ systematically in their propensity to spread. $D_{ij}$ denotes the structural distance between occupations---operationalized in Methods---and $\gamma$ governs how sharply adoption decays with that distance. The framework separates three analytically distinct forces that prior accounts of skill dynamics have not jointly estimated: baseline occupational capacities, skill-specific diffusibility, and relational friction between occupational positions.

\subsection*{Directional asymmetry as the key theoretical extension}

The standard gravity model treats friction as direction-symmetric: $D_{ij}$ penalizes structural distance equally regardless of whether a requirement moves upward or downward in the occupational hierarchy. We argue this constraint is theoretically unjustified. The occupational status hierarchy is not a neutral metric space---it is a structured field in which skill portfolio composition, credentialing infrastructure, and organizational valuation of different capability types vary systematically with position. A requirement moving upward encounters a qualitatively different environment than the same requirement moving downward. Treating both directions as governed by the same friction parameter suppresses the very asymmetry that matters for stratification.

Our central theoretical claim is that friction is directional: resistance to propagation differs depending on whether a requirement moves toward higher- or lower-status destinations. Formally, this means allowing the status gradient between source and target to exert distinct effects depending on its sign---an upward gap generating one friction, a downward gap another. The structural distance term $D_{ij}$ captures proximity-based barriers to diffusion; the directional status gradient captures hierarchy-based barriers. The two are conceptually and empirically separable: a requirement can travel a short structural distance while moving sharply upward in status, or vice versa. Full details on the operationalization of structural distance, the directional status gap, and their empirical separation are provided in Methods.

\subsection*{Why friction should be asymmetric: two structural mechanisms}

The expectation that upward and downward propagation carry distinct friction parameters follows from two structural properties of the labor market. To fix ideas, consider a concrete illustration. Over 2015--2024, nursing occupations increased their requirements for clinical informatics and digital documentation --- socio-cognitive requirements long established in administratively adjacent healthcare roles. The same nursing occupations did not propagate their manual dexterity and physical patient-care requirements upward into those destinations. This contrast is illustrative of the directional logic we formalize below (an interval-censored mini-case is reported in the SI).

First, \textit{directional incorporation asymmetry}: the wage gradient between source and target generates structurally distinct incorporation environments depending on skill type. High-status occupations are organized around socio-cognitive capabilities --- their credentialing infrastructure, organizational routines, valuation environments, and material conditions of work are calibrated to cognitive performance. A socio-cognitive requirement moving upward enters an environment where complementary skills are already present and incorporation is organizationally coherent \parencite{tahlin_skill_2023, alabdulkareem_unpacking_2018}. A sensory/physical requirement making the same upward move encounters the opposite: absent co-specializations, misaligned routines, and a material organization of work in which physical contributions carry low salience, the destination lacks the conditions that would make incorporation functional. The result is not active exclusion but structural indifference. Critically, this mechanism is relational: it is the direction of the wage gradient between source and target that generates distinct incorporation environments, not destination characteristics in isolation.

Second, \textit{structural portability constraint}: a requirement's position in the dependency architecture conditions how freely it can diffuse, independently of destination characteristics. Requirements that anchor long chains of prerequisite capabilities carry co-adoption burdens that grow with dependency depth --- adopting them implies replicating prerequisite stacks and reorganizing task structures around them \parencite{hosseinioun_skill_2025}. Low-nestedness requirements propagate more freely precisely because fewer downstream dependencies bind them to their origin context. This mechanism operates at the level of the requirement itself rather than the dyad.

The opposing directional effects of nestedness documented below emerge from the joint operation of both mechanisms: for socio-cognitive requirements, high dependency depth amplifies upward facilitation because receptive destinations can absorb the prerequisite stack; for sensory/physical requirements, the same depth compounds confinement because unreceptive destinations cannot accommodate it.

Skill domain thus enters through the first mechanism by shaping which incorporation environments are encountered along upward versus downward moves --- including the material and organizational conditions those environments provide or withhold. Nestedness enters through the second by shaping the portability of the requirement itself. These are independent sources of directional friction, not the same constraint counted twice.

\subsection*{A typology of diffusion trajectories}

These two mechanisms jointly imply a structured typology of expected diffusion trajectories. Crossing skill domain with nestedness position yields four theoretical cells, three of which correspond to empirically identifiable trajectories (Table~\ref{tab:typology}). Socio-cognitive scaffolds ---high-nestedness requirements in the cognitive domain---combine receptive destinations with portability constraints that can be absorbed, generating the strongest upward facilitation. Socio-cognitive specialized skills ---low-nestedness cognitive requirements---face neither strong pull nor strong resistance, producing lateral or modestly upward movement. Physical terminal skills ---low-nestedness sensory/physical requirements---encounter structural indifference at high-wage destinations and stronger downward pull, generating the sharpest directional asymmetry. The high-nestedness physical cell is expected to be empirically rare: foundational physical requirements would face simultaneously high structural portability constraint and unreceptive high-status destinations, making sustained upward accumulation unlikely. We refer to this set of predictions — generated jointly by directional incorporation asymmetry and structural portability constraint — collectively as Asymmetric Trajectory Channeling (ATC).

\medskip

\begin{table}[htbp]
\centering

% (opcional) ajusta el ancho si quieres que no se vea tan “chica”
% \begin{minipage}{0.9\linewidth}
\captionsetup{position=bottom,skip=15pt} % <-- caption abajo + separación

\begin{tabular}{lcc}
\toprule
 & \textbf{Low Nestedness} & \textbf{High Nestedness} \\
\midrule
\textbf{Socio-Cognitive} &
    \textit{Specialized} &
    \textit{Scaffolds} \\
 & Lateral or modestly & Strong upward \\
 & upward movement & facilitation \\
\addlinespace
\textbf{Sensory/Physical} &
    \textit{Terminal} &
    $\emptyset$ \\
 & Sharpest downward & Expected to be \\
 & asymmetry & empirically rare \\
\bottomrule
\end{tabular}

\caption{\textbf{Domain--nestedness typology of skill diffusion trajectories.} Cells correspond to combinations of skill domain and nestedness position. Three cells generate empirically expected trajectories; the high-nestedness physical cell ($\emptyset$) is expected to be empirically rare.}
\label{tab:typology}

% \end{minipage}
\end{table}

% ======================
% RESULTS
% ======================
\section*{RESULTS}

\subsection*{Stylized patterns of asymmetric diffusion}

\begin{figure}[!htbp]
  \centering
  \includegraphics[width=\textwidth]{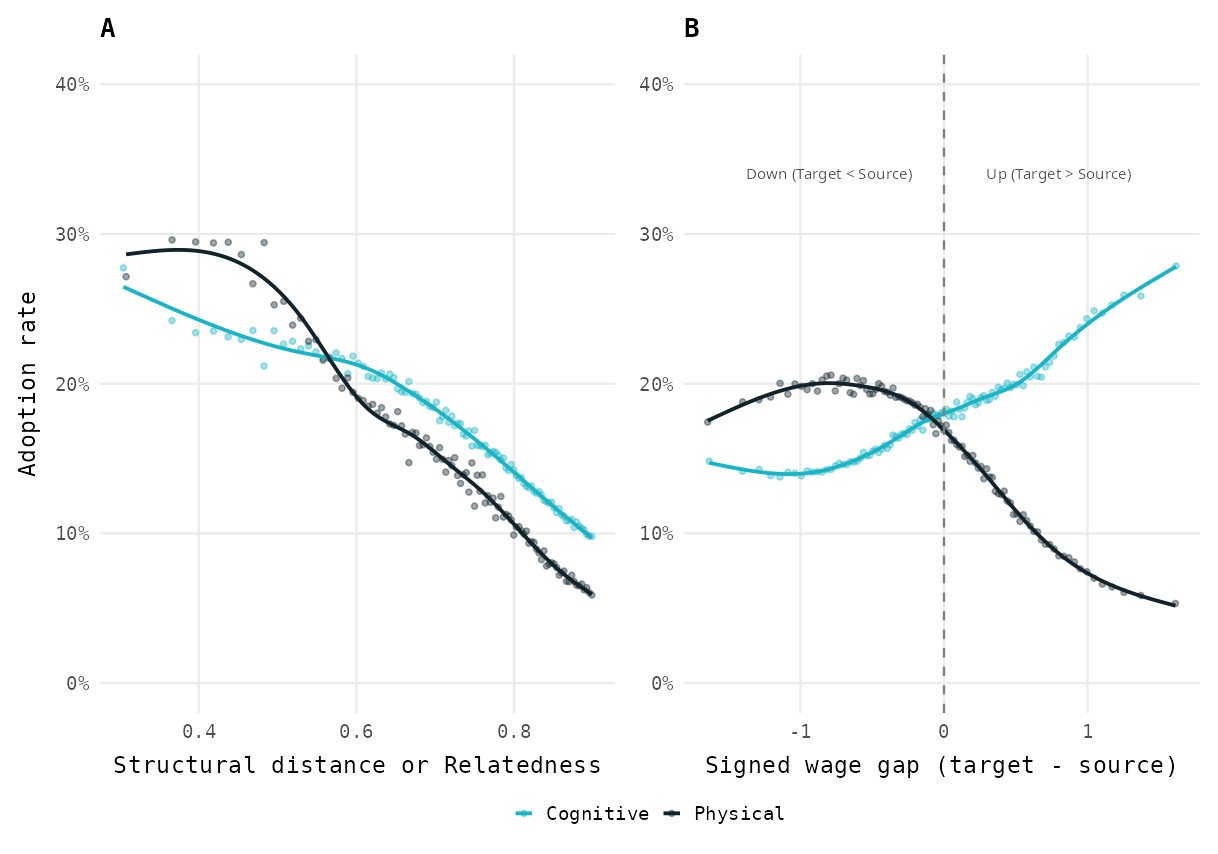}
  \caption{\textbf{Raw adoption rates by structural distance and directional wage gap.}
  Panel A shows adoption rates as a function of occupational relatedness (structural   distance) for socio-cognitive (turquoise) and sensory/physical (black) requirements. Panel B plots adoption rates against the signed wage gap between source and destination occupations; the dashed vertical line separates downward (left) from upward (right) diffusion. Lines are LOWESS smooths; points are binned means across approximately 17.3 million dyadic opportunities (O*NET, 2015--2024).}
  \label{fig:raw_patterns}
\end{figure}

High-wage destinations exhibit markedly different descriptive adoption gradients by skill domain (Figure~\ref{fig:raw_patterns}). Across approximately 17.3 million dyadic diffusion opportunities, socio-cognitive adoption rates rise monotonically as destination wages exceed source wages, consistent with predominantly upward diffusion. For sensory/physical requirements, the pattern reverses: adoption is highest when source and destination are close in status and declines sharply as the upward wage gap widens. Both domains exhibit distance decay, but at different rates: physical requirements lose roughly half their baseline adoption probability over the observed distance range, whereas socio-cognitive requirements decline more gradually and remain comparatively likely to diffuse even at larger structural distances (Panel A).

\begin{figure}[!htbp]
  \centering
  \includegraphics[width=\textwidth]{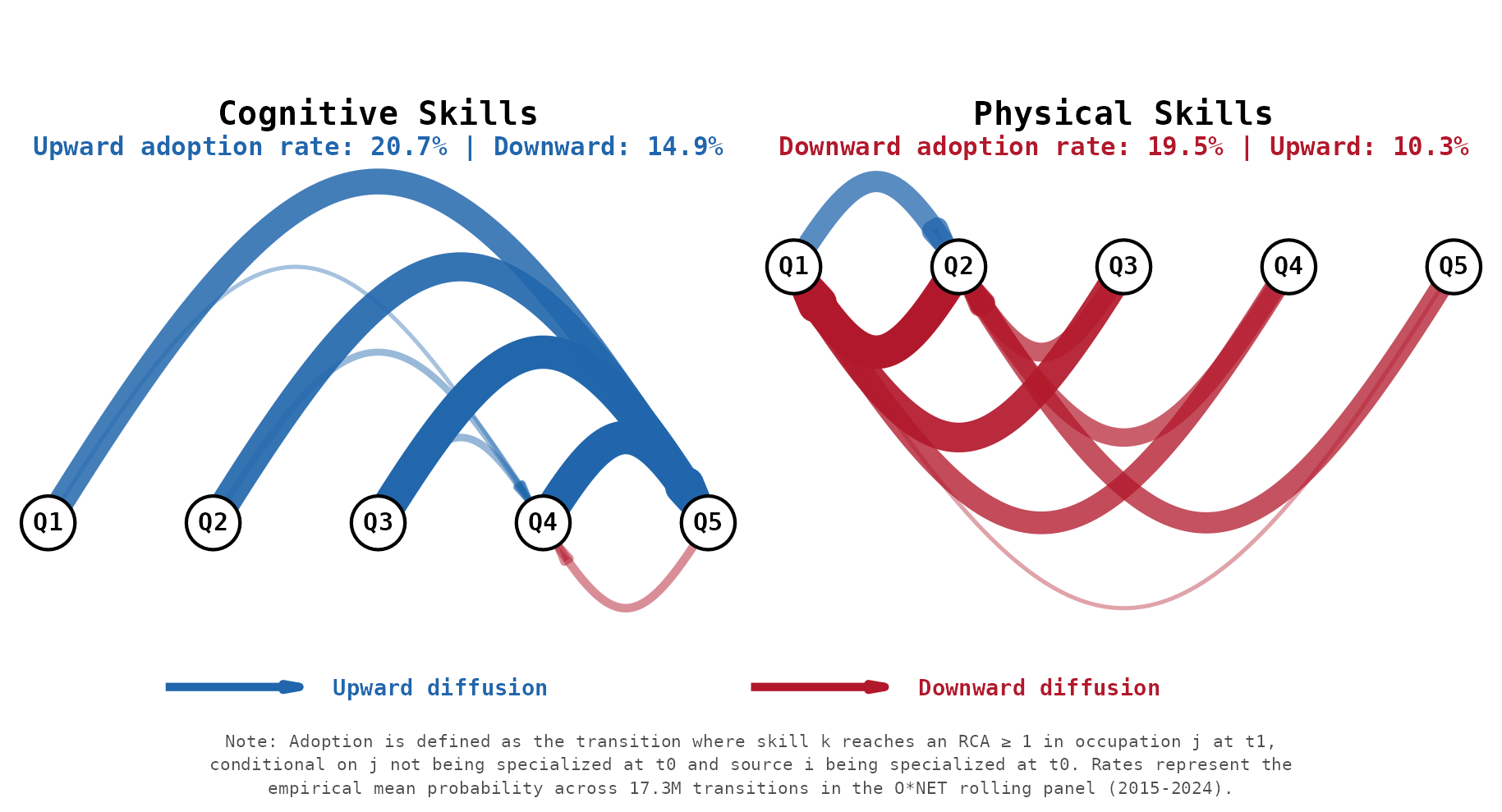}
  \caption{\textbf{Directional skill flows between wage quintiles by domain.}
  Each arc represents the empirical adoption rate from a source quintile to a
  destination quintile, averaged across all skills within each domain
  (O*NET, 2015--2024; $n \approx 17.3$M dyadic opportunities). Blue arcs indicate upward diffusion; red arcs indicate downward diffusion. Arc thickness is proportional to adoption rate. Socio-cognitive requirements (left panel) diffuse upward at 20.7\% versus 14.9\% downward; sensory/physical requirements (right panel) diffuse downward at 19.5\% versus 10.3\% upward. Adoption is defined as the transition where skill $k$ reaches $\mathrm{RCA} \geq 1$ in occupation $j$ at $t_1$, conditional on $j$ not being specialized at $t_0$ and source $i$ being specialized at $t_0$.}
  \label{fig:flow_summary}
\end{figure}

The asymmetry is equally sharp when translated into directional flows between wage quintiles (Figure~\ref{fig:flow_summary}). Socio-cognitive requirements diffuse upward at 20.7\% versus 14.9\% downward --- a gap of nearly six percentage points reflecting systematic, not incidental, upward channeling. Sensory/physical requirements show the mirror image: a downward adoption rate of 19.5\% against an upward rate of only 10.3\%. Cognitive flows concentrate in Q4 and Q5; physical flows concentrate in Q1 and Q2.

Three stylized facts emerge from these raw patterns. \emph{(i)~Directional filtering.} Conditional on distance, socio-cognitive requirements account for excess upward diffusion while sensory/physical requirements concentrate in lateral or downward trajectories --- a gap that persists across all wage quintile pairs. \emph{(ii)~Distance decay asymmetry.} The adoption slope over structural distance is substantially steeper for sensory/physical requirements, indicating that physical skills are more tightly bound to their local occupational context. \emph{(iii)~Boundary constraints.} Cross-domain transfers are suppressed relative to within-domain exchanges, and the penalty is larger for physical requirements attempting to cross into socio-cognitive territory than vice versa. Supplementary Figures~S3--S4 confirm that these domain-level patterns are not driven by a small subset of skills but reflect a pervasive gradient across the full taxonomy. These patterns are robust to alternative distance constructions and RCA threshold sensitivity (Supplementary Table~S1).

\subsection*{Directional status barriers confirm ATC}

\begin{figure}[!htbp]
  \centering
  \includegraphics[width=\textwidth]{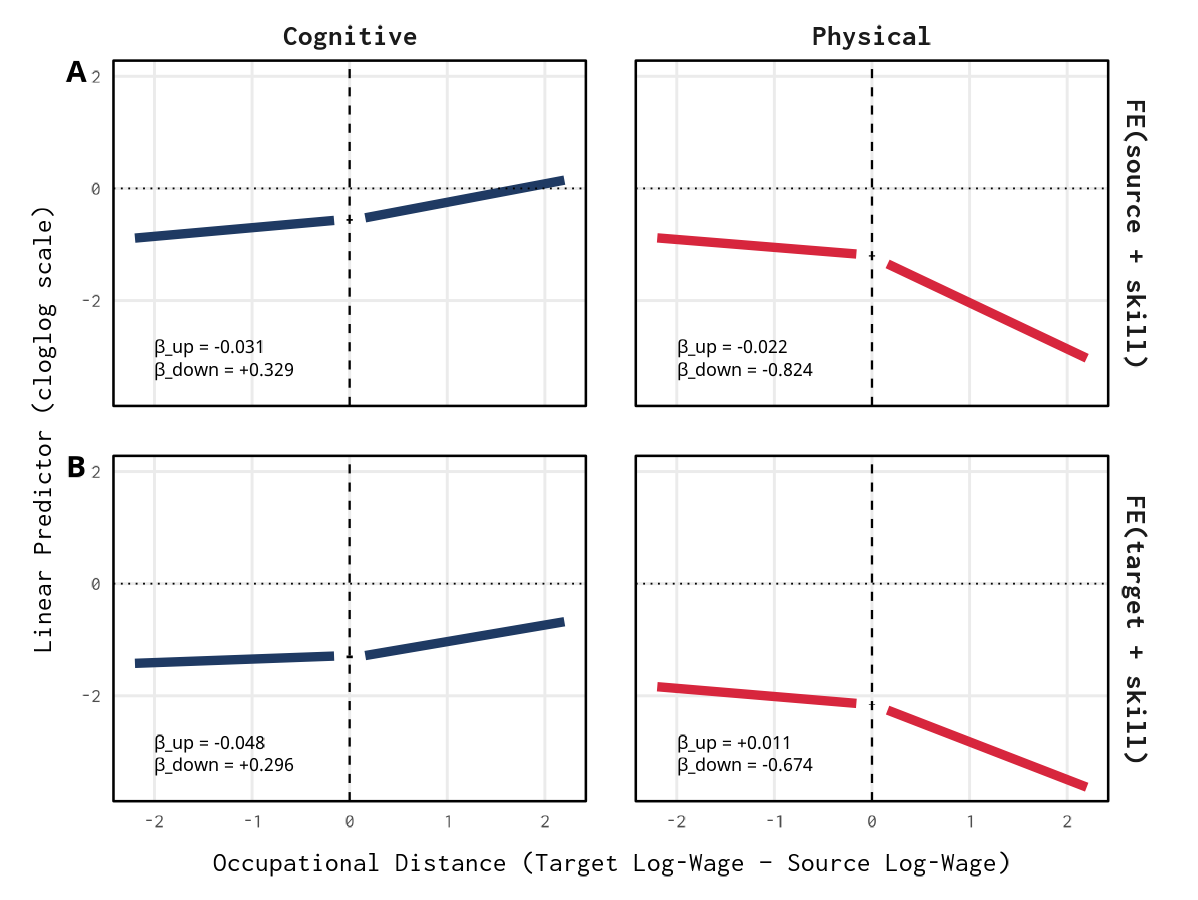}
  \caption{\textbf{Asymmetric trajectory channeling by skill domain.}
Estimated upward ($\beta^{\uparrow}$) and downward ($\beta^{\downarrow}$) friction parameters from the interval-censored gravity model, shown separately for socio-cognitive and sensory/physical requirements. Panel A absorbs source occupation and skill fixed effects; Panel B absorbs target occupation and skill fixed effects. In both specifications, the downward facilitation parameter for sensory/physical requirements ($\beta_{\downarrow} \approx -0.82$ and $-0.67$, respectively; more negative values indicate stronger downward facilitation) substantially exceeds the upward component, which is statistically indistinguishable from zero, while socio-cognitive requirements show comparable friction in both directions. The x-axis plots the signed wage gap (target minus source log-wage); slopes represent directional friction parameters. Points denote point estimates; bars denote 95\% confidence intervals constructed via node-level bootstrap ($B = 1000$ replications).}
  \label{fig:atc_coefficients}
\end{figure}

The gravity estimates reveal a sharp and consistent domain asymmetry in directional friction (Figure~\ref{fig:atc_coefficients}). For socio-cognitive requirements, the upward and downward friction components exhibit a comparatively symmetric profile, indicating no strong directional barrier net of structural distance and occupational heterogeneity. For sensory/physical requirements, the pattern differs qualitatively: the downward component is large in magnitude ($\beta_{\downarrow} \approx -0.82$ in the source fixed-effects specification; more negative implies stronger downward facilitation), whereas the upward component is statistically indistinguishable from zero across specifications. Taken together, the estimates imply pronounced downward channeling for sensory/physical requirements and no evidence of upward facilitation, consistent with the interpretation that high-wage destinations are comparatively unresponsive to additional sensory/physical requirements. This asymmetry is robust across both specifications (Panel A and Panel B). Because socio-cognitive requirements face little upward resistance while sensory/physical requirements face an upward barrier, the evolution of requirements mechanically shifts high-status portfolios toward socio-cognitive complexity and low-status portfolios toward physical terminality.

Distance sensitivity mirrors and reinforces this asymmetry. A one-standard-deviation increase in composite structural distance reduces physical adoption by approximately 70\% more than cognitive adoption --- a differential that persists across alternative distance constructions including skill-network shortest paths and resistance distances (Supplementary Information). Socio-cognitive requirements find receptive environments across a wide range of destinations, bridging large structural gaps; physical requirements remain tightly bound to locally similar occupational contexts.

Two design features strengthen the interpretability of these estimates. First, source and target fixed effects absorb all time-invariant occupational heterogeneity that could otherwise confound directional patterns; the friction parameters are identified from within-occupation variation in propagation direction rather than from cross-sectional differences between occupations. Second, all status gap measures are constructed from 2015 wages and held fixed throughout the window, making them predetermined with respect to 2015--2024 adoption outcomes and reducing concerns about reverse causality. Robustness checks across alternative RCA thresholds, distance constructions, and sub-periods confirm that the directional asymmetry is not an artifact of measurement or model specification (Supplementary Table~S1).

\subsection*{Nestedness amplifies asymmetric channeling}

\begin{figure}[!htbp]
  \centering
  \includegraphics[width=\textwidth]{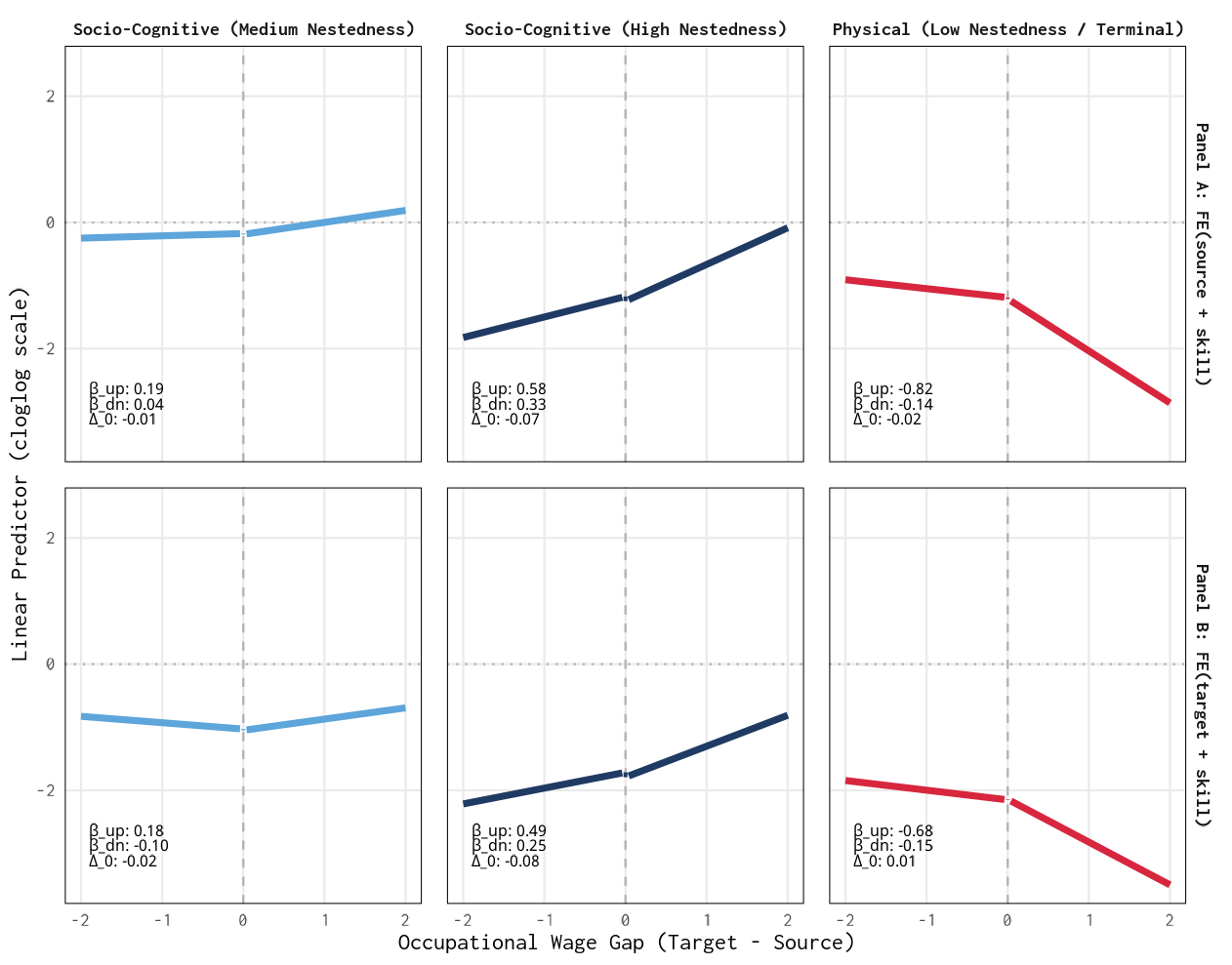}
  \caption{\textbf{Nestedness amplifies asymmetric trajectory channeling.}
Estimated directional friction parameters by domain--nestedness cell, from the interval-censored gravity model. Columns correspond to three empirically distinct trajectory types: SC\_Specialized (low-nestedness socio-cognitive), SC\_Scaffolding (high-nestedness socio-cognitive), and Physical\_Terminal (sensory/physical). Panel A absorbs source occupation and skill fixed effects; Panel B absorbs target occupation and skill fixed effects. For socio-cognitive requirements, upward facilitation increases monotonically with nestedness ($\beta^{\uparrow}$ from 0.19 to 0.58 in Panel A). For sensory/physical requirements, the upward penalty intensifies with nestedness position ($\beta^{\uparrow} = -0.82$ in Panel A), while the high-nestedness physical cell is empirically rare, consistent with the theoretical prediction that foundational physical scaffolds do not accumulate in high-status destinations. The x-axis plots the signed wage gap (target minus source log-wage). Points denote point estimates; bars denote 95\% confidence intervals constructed via node-level bootstrap ($B = 1000$ replications).}
  \label{fig:atc_nestedness}
\end{figure}

Nestedness sharpens the domain asymmetry in predictable and opposing directions (Figure~\ref{fig:atc_nestedness}). For socio-cognitive requirements, higher nestedness systematically reduces upward friction: scaffolding capabilities --- those enabling the broadest range of downstream skills --- travel upward most readily. For sensory/physical requirements, the interaction inverts: low-nestedness physical skills already face upward penalties, but high-nestedness physical skills face penalties that are steeper still. The most structurally embedded physical requirements --- those anchoring the longest dependency chains --- are the most strictly confined to lateral or downward trajectories.

Three empirically distinct trajectories emerge from the domain--nestedness interaction. \textit{Socio-cognitive scaffolds} --- high-nestedness enabling capabilities such as systems analysis and deductive reasoning --- show the strongest upward facilitation ($\beta_{\uparrow} = 0.58$). \textit{Socio-cognitive specialized skills} display a comparatively flat profile, with modest upward facilitation and little directional differentiation. \textit{Physical terminal skills} --- low-nestedness sensory and motor capabilities --- exhibit the sharpest asymmetry: upward moves are strongly penalized ($\beta_{\uparrow} = -0.82$), while the downward component is only mildly facilitative ($\beta_{\downarrow} = -0.14$), yielding net confinement to lateral or downward trajectories. The high-nestedness physical cell is empirically rare, consistent with the theoretical prediction that foundational physical scaffolds do not accumulate in high-status destinations. Nestedness thus does not merely modulate ATC; it specifies the boundary conditions under which upward diffusion is feasible at all.

These three trajectories are stable across time. Splitting the panel into three sub-periods (2015--18, 2019--21, 2022--24) confirms that the domain--nestedness gradient persists in sign and magnitude throughout, with the Physical Terminal asymmetry showing a slight intensification in the most recent period (Supplementary Table~S2). Temporal stability supports the interpretation that directional channeling is a structural property of the skill space --- not a transitory shock --- that reproduces itself as occupational requirements evolve.

% ======================
% DISCUSSION
% ======================
\section*{DISCUSSION}

% 1  

Skill diffusion is not a neutral process. We show that occupational requirements propagate according to a directional rule --- Asymmetric Trajectory Channeling --- that systematically privileges socio-cognitive requirements in upward diffusion while confining sensory/physical requirements to lateral or downward trajectories. This asymmetry is not produced by differences in skill prevalence or aggregate demand: it is identified from within-occupation variation in diffusion direction, net of all time-invariant occupational characteristics. It is sharpened by each skill's position in the nested dependency architecture \parencite{hosseinioun_skill_2025}: scaffolding capabilities travel upward most readily; structurally embedded physical requirements face the steepest confinement. Together, these results establish that occupational stratification is not merely a static sorting outcome --- it reproduces itself through the directional architecture of how skill requirements spread through an already polarized capability space \parencite{alabdulkareem_unpacking_2018}.

% 2

The directional frictions we estimate imply a predictable aggregate consequence: the progressive qualitative depletion of middle-wage destinations. The middle of the wage distribution---routine coordination, clerical, and technical-support occupations that translate procedures into operations and historically bundle basic cognitive monitoring with hands-on execution---is doubly exposed. It loses socio-cognitive requirements that increasingly concentrate in high-wage professional and managerial destinations (Q4--Q5), while absorbing sensory/physical requirements that shift downward into lower-wage frontline work. This is not the routine-task displacement mechanism \parencite{autor_growth_2013, acemoglu_skills_2011}, nor an artifact of offshoring or firm restructuring: it is a complementary propagation mechanism operating through the directional architecture of requirement diffusion, independently of changes in the stock of workers or aggregate demand. This distinction matters acutely in the context of accelerating AI adoption. Recent evidence shows that large language models expose roughly half the U.S. workforce to task-level substitution \parencite{eloundou_gpts_2024}, and that AI-assisted workers achieve substantial productivity gains concentrated in cognitively intensive tasks \parencite{noy_experimental_2023}. If AI-driven productivity gains accrue disproportionately to socio-cognitive work---and our results show that socio-cognitive requirements already face negligible upward resistance---then the propagation rule we document acts as a structural amplifier: technological change does not merely shift demand, it flows through a skill space whose directional architecture channels its benefits upward and its displacements downward. The ATC propagation rule thus provides a structural account of why AI-driven productivity gains need not translate into occupational convergence, even when aggregate upskilling pressure is high \parencite{tong_lower-skilled_2025}.

% 3

These results connect to and extend three converging lines of research. First, relational accounts of stratification locate inequality not in individual attributes but in the structured interactions between positions \parencite{tomaskovic-devey_relational_2019, tahlin_skill_2023}. Our findings extend this tradition by showing that the relational mechanism operates prior to hiring and wage-setting: directional frictions govern which requirements enter an occupation before any worker is selected, embedding hierarchy in the organizational substrate itself. Second, recent work on occupational mobility demonstrates that transitions follow gradational patterns along continuous axes of occupational characteristics, with structural distance predicting mobility smoothly rather than categorically \parencite{york_gradationalism_2025}. The gravity model we estimate is consistent with this gradationalist framework --- but it reveals an asymmetry that gradationalism alone does not capture: distance decay differs fundamentally by direction and domain, producing categorical-like barriers at skill boundaries even within a formally continuous metric. Third, the moderating role of nestedness connects ATC to the economic complexity literature, where capability coherence and structural position in dependency hierarchies predict economic outcomes at the city and country level \parencite{hidalgo_product_2007, hosseinioun_skill_2025}. What ATC adds is the dynamic propagation rule that translates structural position in the skill space into directional friction --- and directional friction into the reproduction of hierarchy over time.

% 4

Our results stand in productive tension with evidence that lower-skilled occupations face the steepest upskilling pressure in employer demand \parencite{tong_lower-skilled_2025}. That work speaks to aspirational upskilling: what organizations signal they want. Our results speak to realized diffusion: which requirements actually propagate across occupations and accumulate in their portfolios. The two accounts are not contradictory --- they illuminate different moments of the same process, and their divergence is precisely where ATC operates. This gap between aspiration and realization acquires new urgency in the context of AI-driven task substitution. Machine learning capabilities are distributed unevenly across the skill space \parencite{brynjolfsson_what_2017}, with the highest substitution potential concentrated in well-specified physical and routine cognitive tasks \parencite{acemoglu_robots_2020} --- the same requirements our estimates show are most structurally confined. If automation pressure is highest precisely where upward propagation is most structurally blocked, then displaced physical workers face a compounded barrier: the requirements they hold are both most exposed to substitution and least able to traverse the domain boundary that separates them from cognitively intensive destinations. The ATC propagation rule thus identifies a structural mechanism by which technological disruption and skill-space architecture jointly reproduce occupational immobility, independently of any individual worker's retraining effort or employer's hiring intent.

% 5

The propagation rule we document is not invariant --- its strength is conditioned by the institutional and organizational structure of labor markets. ATC gradients should be weakest where credentialing is fungible, occupational boundaries are permeable, and skill complementarities are modular; they should be strongest in credential-intensive, compliance-heavy, or safety-critical domains where switching costs are high and valuation regimes actively police domain boundaries \parencite{tahlin_skill_2023}. Urban and organizational environments that increase job connectivity \parencite{moro_universal_2021} should attenuate the distance component of ATC by multiplying effective exposure opportunities, while labor market fragmentation \parencite{lin_network_2022} should steepen it. These boundary conditions point toward two concrete intervention levers. The first is upward penalty reduction through domain bundling: pairing physical requirements with socio-cognitive complements --- safety compliance with digital inspection protocols, manual precision with quality analytics, physical coordination with systems monitoring --- can shift the effective domain classification of a requirement bundle sufficiently to reduce upward friction. The second is effective distance compression through credential bridges: stackable credentials and cross-functional rotations that bind complementary requirements convert latent structural proximity into realized adoption opportunities. These two levers interact: friction reduction yields larger gains where exposure density is already high \parencite{moro_universal_2021}. Policy design should therefore prioritize requirements with high dependency reach but strong distance sensitivity --- those with the largest gap between structural potential and realized propagation --- as these represent the highest-return targets for intervention.

% 6

Several limitations bound the scope of these conclusions. ATC characterizes the propagation rule of skill requirements at the occupational level; our analysis does not reveal the organizational decision processes that produce individual adoptions, nor does it speak to the general equilibrium dynamics that accompany shifting skill demands at the market level \parencite{alabdulkareem_unpacking_2018}. How exogenous shocks to the capability space --- new technologies, licensing reforms, or credentialing changes --- alter the propagation rule itself remains an open question. The analysis is grounded in U.S. occupational data, and the ATC gradients we document reflect institutional conditions --- credentialing regimes, wage-setting structures, union density --- specific to the U.S. labor market. We expect the propagation rule to operate wherever polarized and nested skill architectures are present, conditions documented across labor markets \parencite{hosseinioun_skill_2025}, but cross-national tests exploiting variation in institutional context are needed to identify which components of ATC are structural universals and which are institutionally contingent. Empirically, O*NET's rolling panel design introduces measurement latency that limits causal interpretation of annual timing and short-run hazard dynamics, and our nestedness measures are inferred from co-specialization patterns rather than from exogenous variation in curricular or licensing requirements. We cannot directly observe inter-occupational transmission events; the propagation interpretation is supported by the conditional structure of the risk set and reinforced by the robustness of directional asymmetries to source-identity permutation. Establishment-level data on internal mobility ladders, job posting corpora, and procurement shocks would allow direct tests of the organizational mechanisms through which directional friction is enacted.

% 7

The labor market's skill hierarchy persists not because occupational requirements are static, but because the process by which they change is itself directionally structured. A simple propagation rule --- downward facilitation and upward indifference for sensory/physical requirements, cross-gradient permeability for socio-cognitive ones, both amplified by position in the nested dependency architecture --- connects the micro-level decisions of organizations to the macro-level reproduction of stratification. Critically, this rule requires no assortative preferences, discriminatory intent, or coordinated action: just as assortative social outcomes can emerge from structural constraints without assortative preference \parencite{xie_assortative_2015}, selective skill channeling emerges from the structural properties of the capability space itself. By specifying where change is most and least likely to travel, ATC identifies the precise points in the propagation process where intervention can open new upgrading paths --- and explains why, absent such intervention, occupational hierarchy reproduces itself silently, through the architecture of the space rather than the intentions of its inhabitants.

% ======================
% METHODS
% ======================

\section*{METHODS}

\subsection*{Data and endpoint design}

We draw occupation-level skill requirement data from O*NET across two extracts anchored at baseline ($t_0 = 2015$) and endline ($t_1 = 2024$). O*NET reports the importance and required level of 161 skill, knowledge, and ability items for approximately 873 standardized occupations via a rolling update design in which a subset of occupations is refreshed each year. We use an endpoint window rather than annual panels to match this structure: because only a fraction of occupations is resurveyed in any given year, intermediate-wave non-change reflects measurement timing as much as substantive stability. The nine-year window is long enough that all occupations receive at least one full refresh within the horizon, allowing adoption to be measured cleanly at endline without imputing annual timing. Details on O*NET extraction, harmonization, and refresh scheduling are available from the authors upon request and will be released upon publication.
Occupational wages are drawn from the Bureau of Labor Statistics Occupational Employment and Wage Statistics (OEWS). All wage measures are constructed from 2015 values and held fixed throughout the observation window --- predetermined with respect to adoption outcomes --- to eliminate reverse causality from high-adoption skills mechanically attracting higher wages. Throughout, 'diffusion' denotes directional updating of occupational requirements under a risk-set design that requires a plausible source; we make no claim about direct transmission events between specific occupational pairs.

\subsection*{Diffusion opportunities and adoption outcome}

The unit of analysis is a directed triple $(i, j, s)$: source occupation $i$, target occupation $j$, and skill $s$. A triple enters the risk set if $i$ is specialized in $s$ at baseline and $j$ is not, constituting a diffusion opportunity --- a configuration in which $s$ can plausibly propagate from $i$ to $j$. Applying this rule across all ordered occupation pairs and the full 161-skill taxonomy yields approximately 17.3 million diffusion opportunities. Skill specialization is defined via Revealed Comparative Advantage \parencite{alabdulkareem_unpacking_2018}: occupation $j$ is specialized in $s$ if $\mathrm{RCA}(j,s) \geq 1$, where RCA compares the relative importance of $s$ in $j$ to its economy-wide average. RCA normalizes for skill ubiquity, 
identifying specialization relative to the aggregate distribution of requirements rather than raw importance scores. The binary outcome $Y_{ijs} = 1$ if $j$ crosses the specialization threshold by 2024, conditional on not being specialized at 2015. Full RCA derivation and threshold sensitivity checks are available from the authors upon request and will be released upon publication.

\subsection*{Status directionality and structural distance}

The ATC hypothesis requires separating two analytically distinct sources of friction: movement along the occupational status hierarchy, and barriers due to task dissimilarity. We operationalize these separately.

\textbf{Status directionality.} Using predetermined 2015 log-wages, we define the signed status gap $G_{ij} = \log\bar{W}_j - \log\bar{W}_i$ and decompose it into directional components:
\[
\Delta^{\uparrow}_{ij} = \max(0,\, G_{ij}), \qquad 
\Delta^{\downarrow}_{ij} = \min(0,\, G_{ij}),
\]
allowing upward and downward propagation to be governed by distinct friction parameters without imposing symmetry across directions. A direction-entry indicator $\mathbbm{1}[G_{ij}>0]$ additionally captures discrete threshold effects at the status boundary independently of gap magnitude. Because $\Delta^{\downarrow}_{ij}\le 0$ by construction, more negative estimates of $\Theta^{\downarrow}$ imply stronger facilitation along downward trajectories.

\textbf{Structural distance.} We measure task relatedness between source and target as one minus the cosine similarity of their O*NET skill vectors at baseline. This captures portfolio proximity independently of status, ensuring that directional terms index movement along the wage hierarchy rather than proxying for task similarity. Distance is computed from 2015 skill vectors and is therefore predetermined with respect to adoption outcomes. Robustness checks replace cosine distance with skill-network shortest-path and resistance distances; conclusions are unchanged (Supplementary Table~S1).

\subsection*{Gravity model}

We model skill adoption over the 2015--2024 window using a gravity-type specification with a complementary log-log (cloglog) link, appropriate for a binary outcome with a proportional-hazards interpretation over a discrete interval. Because adoption is observed only at the end of the window, the specification can be read as a discrete-time hazard integrated over 2015--2024 under interval observation.

Let the directed wage gap between occupations be
\[
\Delta_{ij} \equiv \log w_j - \log w_i,
\]
and define the upward and downward components
\[
\Delta^{\uparrow}_{ij} \equiv \max(0,\Delta_{ij}), 
\qquad
\Delta^{\downarrow}_{ij} \equiv \min(0,\Delta_{ij}),
\]
together with an indicator for crossing the status boundary,
\[
\mathbbm{1}[\Delta_{ij}>0].
\]
The gravity-hazard benchmark linear predictor for triple $(i, j, s)$ is:
\begin{equation}
\eta_{ijs} \;=\; \alpha_i + \alpha_j + \alpha_s
\;+\; \Theta^{\uparrow}(g)\,\Delta^{\uparrow}_{ij}
\;+\; \Theta^{\downarrow}(g)\,\Delta^{\downarrow}_{ij}
\;+\; \kappa(g)\,\mathbbm{1}[\Delta_{ij}>0]
\;+\; \delta(g)\,\mathrm{dist}_{ij},
\label{eq:gravity_main}
\end{equation}

\noindent where $\alpha_i$, $\alpha_j$, and $\alpha_s$ are source, target, and skill fixed effects; $\Theta^{\uparrow}(g)$ and $\Theta^{\downarrow}(g)$ are domain- or archetype-specific directional friction parameters; $\kappa(g)$ captures a discrete barrier at the status boundary; and $\delta(g)$ is a domain-specific distance decay coefficient. ATC predicts $\Theta^{\uparrow}$ and $\Theta^{\downarrow}$ to differ systematically by domain and by nestedness position.

\noindent In practice, we implement $g$ via interaction with the full set of channel terms,
$(\Delta^{\uparrow}_{ij},\Delta^{\downarrow}_{ij},\mathbbm{1}[\Delta_{ij}>0],\mathrm{dist}_{ij})\times g$,
so that each archetype/domain has its own slope and discontinuity parameters.

\noindent \textbf{Benchmark and estimands.} Eq.~\ref{eq:gravity_main} summarizes the saturated gravity-hazard benchmark. Because the directional gap terms are constructed directly from occupation wage levels, simultaneously absorbing source and target fixed effects renders the directional parameters $(\Theta^{\uparrow}, \Theta^{\downarrow}, \kappa)$ uninformative. We therefore report two complementary fixed-effect decompositions that recover destination-side and origin-side channeling estimands while maintaining a common structural interpretation of the gravity channel. The estimated $\Theta^{\uparrow}$ and $\Theta^{\downarrow}$ are channel parameters: they quantify directional frictions that operate net of origin/destination heterogeneity and structural distance.

\subsection*{Complementary estimands via fixed-effect decomposition}

Following \textcite{lundberg_what_2021}, we emphasize that alternative fixed-effect structures correspond to distinct target quantities. The ATC propagation rule operates on both sides of the dyad: destination occupations may be selectively receptive, and source occupations may be differentially emissive. To preserve identification of directional gaps while accounting for persistent heterogeneity, we estimate two complementary specifications.

\emph{Panel A --- destination-side estimand.} Source occupation and skill fixed effects ($\alpha_i$, $\alpha_s$) absorb time-invariant emissive heterogeneity at the origin. The directional friction parameters are identified from differences in how the \emph{same source} diffuses toward destinations with varying wage gaps. This estimand captures destination-side receptivity to directional channeling net of origin characteristics. Estimation absorbs the fixed effects in a high-dimensional GLM framework under the cloglog link, and standard errors are clustered by source occupation, target occupation, and skill.

\emph{Panel B --- origin-side estimand.} Target occupation and skill fixed effects ($\alpha_j$, $\alpha_s$) absorb time-invariant absorptive heterogeneity at the destination. The directional friction parameters are identified from differences in how sources with varying wage gaps propagate toward the \emph{same destination}. This estimand captures origin-side emissive constraints net of destination characteristics. Standard errors are clustered by source occupation, target occupation, and skill.

Concordant directional asymmetries across both estimands support interpreting ATC as a property of the diffusion channel itself rather than of either endpoint in isolation. Technical details of the identification argument and sensitivity to alternative fixed-effect structures are available from the authors upon request and will be released upon publication.

\subsection*{Nestedness and ATC archetypes}

To test whether position in the dependency architecture moderates directional friction, we characterize each skill by its contribution to system-wide nestedness, $c_s$, computed as the standardized (z-scored) change in NODF \parencite{hosseinioun_skill_2025} when skill $s$ is removed from the baseline occupation--skill matrix. Both $c_s$ and domain classification are constructed from 2015 data and are therefore predetermined with respect to adoption outcomes.

We classify skills into three empirically motivated archetypes by crossing domain with nestedness position: \textit{SC\_Scaffolding} (socio-cognitive, $c_s \geq$ median within domain), \textit{SC\_Specialized} (socio-cognitive, $c_s <$ median), and \textit{Physical\_Terminal} (sensory/physical, all nestedness levels). The high-nestedness physical cell is empirically rare and treated as a boundary case, consistent with the theoretical prediction that structurally embedded physical requirements do not accumulate in high-status destinations; we report it separately in Supplementary Methods and show results are unchanged when retaining it as its own category. Archetype-specific friction parameters are estimated by interacting archetype indicators with $\Delta^{\uparrow}$, $\Delta^{\downarrow}$, $\mathbbm{1}[G_{ij}>0]$, and $\mathrm{dist}_{ij}$ in both Panel A and Panel B. Full derivation of $c_s$ and NODF construction are available from the authors upon request and will be released upon publication.

\subsection*{Inference and robustness}

Standard errors are clustered three-way by source occupation, target occupation, and skill to account for network dependence in the triadic data structure. Confidence intervals reported in figures are verified via node-level bootstrap ($B = 1{,}000$ replications) resampling occupations rather than dyads, preserving the dependence structure of the network. Four robustness checks confirm that directional asymmetries are not artifacts of measurement or specification: (i)~threshold placebo tests displacing the status-boundary cutoff away from $G_{ij} = 0$; (ii)~domain-label permutation tests collapsing the skill-domain alignment while preserving risk-set structure; (iii)~within-stratum permutation tests severing source identity from adoption; and (iv)~RCA threshold sensitivity across cutpoints 0.9, 1.0, and 1.1 and across sub-periods (Supplementary Table~S1). All estimation uses \texttt{fixest} in R \parencite{berge_fixest_2025}.

% ======================
% ADDITIONAL INFORMATION 
% ======================
\section*{Data availability}
All derived data needed to evaluate the conclusions are available from the corresponding author upon reasonable request. O*NET releases (2015–2024), BLS occupational wages, and NAICS concordances are publicly available; access instructions are provided in the Supplementary Information.

\section*{Code availability}
All analysis code (data processing, model estimation, figure generation, and replication scripts) will be made publicly available under an open-source license in a versioned repository with a reproducible environment file (R lockfile/\texttt{renv} and container spec). A pre-release link is available from the corresponding author upon request.


\clearpage
\begin{thebibliography}{10}

\bibitem{acemoglu_skills_2011}
D.~Acemoglu, D.~Autor, {\it Handbook of {Labor} {Economics}\/} (Elsevier, 2011), vol.~4, pp. 1043--1171.

\bibitem{autor_skill_2003}
D.~H. Autor, F.~Levy, R.~J. Murnane, The {Skill} {Content} of {Recent} {Technological} {Change}: {An} {Empirical} {Exploration}.
\newblock {\it The Quarterly Journal of Economics\/} {\bf 118}, 1279 (2003).

\bibitem{alabdulkareem_unpacking_2018}
A.~Alabdulkareem, {\it et~al.\/}, Unpacking the polarization of workplace skills.
\newblock {\it Science Advances\/} {\bf 4}, eaao6030 (2018).

\bibitem{hosseinioun_skill_2025}
M.~Hosseinioun, F.~Neffke, L.~Zhang, H.~Youn, Skill dependencies uncover nested human capital.
\newblock {\it Nature Human Behaviour\/}  (2025).

\bibitem{del_rio-chanona_occupational_2021}
R.~M. Del Rio-Chanona, P.~Mealy, M.~Beguerisse-Díaz, F.~Lafond, J.~D. Farmer, Occupational mobility and automation: a data-driven network model.
\newblock {\it Journal of The Royal Society Interface\/} {\bf 18}, 20200898 (2021).

\bibitem{frank_small_2018}
M.~R. Frank, L.~Sun, M.~Cebrian, H.~Youn, I.~Rahwan, Small cities face greater impact from automation.
\newblock {\it Journal of The Royal Society Interface\/} {\bf 15}, 20170946 (2018).

\bibitem{mealy_interpreting_2019}
P.~Mealy, J.~D. Farmer, A.~Teytelboym, Interpreting economic complexity.
\newblock {\it Science Advances\/} {\bf 5}, eaau1705 (2019).

\bibitem{tinbergen_shaping_1962}
J.~Tinbergen, Shaping the world economy; suggestions for an international economic policy.  (1962).

\bibitem{hidalgo_product_2007}
C.~A. Hidalgo, B.~Klinger, A.-L. Barabási, R.~Hausmann, The product space conditions the development of nations.
\newblock {\it Science\/} {\bf 317}, 482 (2007). Publisher: American Association for the Advancement of Science.

\bibitem{tahlin_skill_2023}
D.~Avent-Holt, D.~Tomaskovic-Devey, {\it A {Research} {Agenda} for {Skills} and {Inequality}\/}, M.~Tåhlin, ed. (Edward Elgar Publishing, 2023), pp. 217--232.

\bibitem{autor_growth_2013}
D.~H. Autor, D.~Dorn, The {Growth} of {Low}-{Skill} {Service} {Jobs} and the {Polarization} of the {US} {Labor} {Market}.
\newblock {\it American Economic Review\/} {\bf 103}, 1553 (2013).

\bibitem{eloundou_gpts_2024}
T.~Eloundou, S.~Manning, P.~Mishkin, D.~Rock, {GPTs} are {GPTs}: {Labor} market impact potential of {LLMs}.
\newblock {\it Science\/} {\bf 384}, 1306 (2024).

\bibitem{noy_experimental_2023}
S.~Noy, W.~Zhang, Experimental evidence on the productivity effects of generative artificial intelligence.
\newblock {\it Science\/} {\bf 381}, 187 (2023).

\bibitem{tong_lower-skilled_2025}
D.~Tong, L.~Wu, J.~A. Evans, Lower-skilled occupations face greater upskilling pressure in {U}.{S}. job ads.
\newblock {\it Nature Communications\/} {\bf 17}, 1237 (2025).

\bibitem{tomaskovic-devey_relational_2019}
D.~Tomaskovic-Devey, D.~R. Avent-Holt, {\it Relational inequalities: an organizational approach\/} (Oxford University Press, New York, 2019).

\bibitem{york_gradationalism_2025}
H.~York, X.~Song, Y.~Xie, Gradationalism {Revisited}: {Intergenerational} {Occupational} {Mobility} {Along} {Axes} of {Occupational} {Characteristics}.
\newblock {\it American Journal of Sociology\/} {\bf 130}, 976 (2025).

\bibitem{brynjolfsson_what_2017}
E.~Brynjolfsson, T.~Mitchell, What can machine learning do? {Workforce} implications.
\newblock {\it Science\/} {\bf 358}, 1530 (2017). Publisher: American Association for the Advancement of Science.

\bibitem{acemoglu_robots_2020}
D.~Acemoglu, P.~Restrepo, Robots and {Jobs}: {Evidence} from {US} {Labor} {Markets}.
\newblock {\it Journal of Political Economy\/} {\bf 128}, 2188 (2020).

\bibitem{moro_universal_2021}
E.~Moro, {\it et~al.\/}, Universal resilience patterns in labor markets.
\newblock {\it Nature Communications\/} {\bf 12}, 1972 (2021).

\bibitem{lin_network_2022}
K.-H. Lin, K.~Hung, The {Network} {Structure} of {Occupations}: {Fragmentation}, {Differentiation}, and {Contagion}.
\newblock {\it American Journal of Sociology\/} {\bf 127}, 1551 (2022).

\bibitem{xie_assortative_2015}
Y.~Xie, S.~Cheng, X.~Zhou, Assortative mating without assortative preference.
\newblock {\it Proceedings of the National Academy of Sciences\/} {\bf 112}, 5974 (2015).

\bibitem{lundberg_what_2021}
I.~Lundberg, R.~Johnson, B.~M. Stewart, What {Is} {Your} {Estimand}? {Defining} the {Target} {Quantity} {Connects} {Statistical} {Evidence} to {Theory}.
\newblock {\it American Sociological Review\/} {\bf 86}, 532 (2021).

\bibitem{berge_fixest_2025}
L.~Berge, S.~Krantz, G.~McDermott, R.~Lenth, K.~Butts, fixest: {Fast} {Fixed}-{Effects} {Estimations}. (2025).

\end{thebibliography}
\end{document}